\documentstyle[aps,preprint]{revtex}
\begin{document}
\draft
\title {SEARCH FOR EXTREMELY DEFORMED FISSION FRAGMENT IN $^{28}$Si+$^{28}$Si
WITH EUROGAM PHASE II} 

\author { C. Beck, R. Nouicer, N. Aissaoui, T. Bellot, G. de France,
D. Disdier, G. Duch\`ene, \\
A. Elanique, R.M. Freeman, F. Haas, \\ 
A. Hachem, F. Hoellinger, D. Mahboub, D. Pr\'evost, V. Rauch } 

\address{\it Institut de Recherches Subatomiques, Institut National de Physique
Nucl\'eaire et de Physique des Particules - Centre National de la Recherche
Scientifique/Universit\'e Louis Pasteur,B.P.28, F-67037 Strasbourg Cedex 2,
France } 

\author { S.J. Sanders, T. Catterson, A. Dummer, F.W. Prosser }

\address {\it Department of Physics and Astronomy, University of Kansas,
Lawrence, USA }

\author { A. Szanto de Toledo }

\address {\it Depatamento de Fisica Nuclear, University of Sa\~o Paulo, Brazil }

\author { Sl. Cavallaro}

\address {\it Dipartimento di Fisica dell'Universit\'a di Catania, INFN and 
LNS Catania, I-95129 Catania, Italy }

\date{\today}
\maketitle

\newpage

\begin{abstract}
{A high-resolution study of fragment-fragment-$\gamma$ triple coincident
measurements of the symmetric-mass fission exit-channel from the
$^{28}$Si+$^{28}$Si reaction has been performed at the VIVITRON Tandem facility
by using the EUROGAM Phase II $\gamma$-ray spectrometer. The bombarding energy
E$_{lab}$($^{28}$Si) = 111.6 MeV has been chosen to populate a well known
quasi-molecular resonance in $^{56}$Ni. Evidence is presented for a selective
population of states in $^{28}$Si fragments arising from the symmetric-fission
of the $^{56}$Ni compound nucleus. In the resonant region the enhanced
population of the K$^{\pi}$ = 3$_{1}^{-}$ band of the $^{28}$Si nucleus,
indicative of an oblate deformed shape, appears to play a significant role in
the collision processes. The resonant behavior in the elastic and inelatic
$^{28}$Si+$^{28}$Si exit-channels is found to be correlated to strong
disalignment features of the di-nuclear oblate-oblate system with an
equator-equator stable configuration. } 
\end{abstract} 

\newpage

\centerline {\bf I. Introduction }

\vskip 1.0 cm

In recent years, extensive efforts have been made in the study of the
fusion-fission (FF) dynamics of very light di-nuclear systems
\cite{Fa96,Be96,No96,Ma97} (40 $\leq$ A$_{cn}$ $\leq$ 60). Statistical decay
processes \cite{Fa96,Ma97} have been demonstrated to influence reaction
channels that have been previously explored in terms of heavy-ion resonance
mechanisms \cite{Be81a}. The FF mechanim is known to play a significant role at
spins slightly above the grazing angular momentum; the nuclear configuration
leading to the resonance behavior is only slightly more extended than that of
the nuclear saddle point. According to the number of open channels (NOC) model
\cite{Be95}, it has been shown that the coexistence of statistical fission from
the $^{56}$Ni and the resonances arising from very deformed configurations of
this composite system are well explained. The NOC model \cite{Be94} predicts
that the $^{28}$Si+$^{28}$Si collision has all the required features associated
with surface transparency that are generally observed in much lighter
``resonating" systems such as $^{12}$C+$^{12}$C and $^{16}$O+$^{16}$O. 

It has been suggested that a coherent framework may exist which connects the
topics of heavy-ion quasi-molecular resonances, superdeformation effects and
fission shape isomerism \cite{Be95,Be94,Be85}. In particular shell-stabilized,
highly deformed configurations (as illustrated by the existence of
``superdeformed" second minima in potential energy surfaces as calculated by
the Nilsson-Strutinsly model shown in fig.1) in the $^{56}$Ni compond system
have been conjectured \cite{Be85} to explain the strong resonant behavior as
observed at large angles for the $^{28}$Si+$^{28}$Si reaction \cite{Be81a}.
Very narrow widths of about 150 keV high-spin resonances (see fig.2) are found
to be well correlated in the elastic and inelastic decay channels of single and
mutual excitations. This narrow and regular structure appears to correspond to
a complex pattern of isolated high-spin resonances in the di-nuclear system
with angular momenta, obtained from elastic scattering angular distributions
\cite{Be81b}, ranging from 34$\hbar$ to 42$\hbar$ following the grazing partial
wave sequence. 

This very striking quasi-molecular resonant structure can possibly be
correlated to a rather special subset of high-spin states stabilized against
mixing into the more numerous compound nucleus states by some special symmetry
\cite{Be85}. As a matter of fact recent theoretical investigations have
indicated that shell-stabilized ``hyperdeformed" shapes may exist in the
$^{56}$Ni nucleus with large angular momenta \cite{Ra92,Ue94}. By the use of a
molecular model proposed by Uegaki and Abe \cite{Ue89,Ue93}, a stable
configuration of the di-nuclear system is found to be an equator-equator
touching one, due to the oblate deformed shape of the $^{28}$Si nuclei
\cite{Ue94}. In this report, we present for the first time experimental results
using very powerful coincidence techniques which indicate the possible
occurence of a butterfly mode excitation responsible of the quasi-molecular
resonant structure. 

\newpage

\centerline{\bf II. Experimental Techniques } 

\vskip 1.0 cm

The aim of the experiment we have performed at the VIVITRON tandem facility of
IReS Strasbourg with the EUROGAM Phase II multi-detector array was to search
for highly deformed bands in the $^{28}$Si nucleus as produced in the
$^{28}$Si+$^{28}$Si FF reaction. 

A bombarding energy E$_{lab}$($^{28}$Si) = 111.6 MeV has been chosen to
populate the well known 38$^{+}$ resonance \cite{Be81a,Be81b,Be85} as shown in
fig.2. The experiment has been carried out in triple coincidence modes 
(fragment-fragment-$\gamma$) with two fission fragments mass identified with
two pairs of large-area position-sensitive Si(surface-barrier) detectors
placed on either side of the beam axis by using standard kinematic coincidence
techniques \cite{Be81c} and the $\gamma$-rays detected in EUROGAM Phase II
multi-detector array. To establish the absolute normalization of the binary
products yields, the elastic scattering data as measured by Betts \cite{Be97}
at a bombarding energy E$_{lab}$($^{28}$Si) = 112 MeV have been used. Energy
and relative efficiency calibrations of EUROGAM Phase II were obtained with
standard $\gamma$-ray sources and a AmBe source for the higher energy
$\gamma$-ray region \cite{No97}. 

\newpage

\centerline{\bf III. Fragment-fragment coincidence data }

\vskip 1.0 cm

Fig.3 shows the gated two-dimensional spectrum of the ejectile energy E$_{3}$
as a function of the calculated excitation energy E$_{x}$ for the
$^{28}$Si+$^{28}$Si symmetric-fission exit-channel. The vertical band
correspond to different excited states in the two $^{28}$Si fragments, whereas
the regular concentrations of yields are due to strongly structured angular
distributions. 

The excitation-energy spectrum of the $^{28}$Si+$^{28}$Si symmetric-fission
exit-channel, displayed in fig.4, exhibits a very striking and well structured
behaviour up to high excitation energy (E$_{x}$ $\approx$ 15 MeV). The low
excitation-energy peaks correspond essentially to the elastic scattering and
inelastic scattering to low lying states of $^{28}$Si (mainly the 1.78 MeV
2$^{+}$, and 4.62 MeV 4$^{+}$ states), strong peaks are observed at much more
negative Q-values and more likely arise from mutual scatterings. The states
assignments are indicative since the higher E$_{x}$ lying peaks are speculative
although the locations of these peaks are consistent with mutual excitation of
yrast states in both fragments as suggested previously \cite{Be81c}. The high
selectivity in the final-state population will be discussed more in detail in
the analysis of the $\gamma$-ray coincident results. 

The angular distributions are extracted from the projections on the E$_{3}$
axis with excitation energy gates as defined by the states of fig.4. The
$^{28}$Si($^{28}$Si,$^{28}$Si)$^{28}$Si identical particle exit-channels at
E$_{lab}$ = 111.6 MeV are found to have at backward angles (between 70$^{o}$
$\leq$ $\theta$$_{c.m.}$ $\leq$ 110$^{o}$) strongly oscillatory angular
distributions in the elastic, inelastic and mutual excitation channels as shown
in fig.5 (points). The elastic cross sections observed in this angular region
are from one to three orders of magnitude (near $\theta$$_{c.m.}$ = 90$^{o}$)
larger than those obtained from optical-model calculations using standard
strongly absorbing potential parameters which give a good account of the data
forward of $\theta$$_{c.m.}$ = 70$^{o}$. This behavior suggests the presence of
a quite longer time-scale mechanism than the direct and multistep processes
that dominate at more forward angles ($\theta$$_{c.m.}$ $\leq$ 70$^{o}$). In
this latter case the scattering processes have a characteristic dependence of
the cross sections (with a steep and smooth fall off beyond the grazing angle
$\theta$$_{c.m.}$ $\approx$ 44$^{o}$ as deduced from the elastic scattering
\cite{Di81}) under the influence of Coulomb repulsion with strong absorption
for small impact parameters. 

The present large-angle high-quality data, with good angular (position)
resolution and high statistics, are well described by the curves of fig.5 as
calculated by P$_{L}^{2}$(cos$\theta$$_{c.m.}$) shapes with L = 38$\hbar$ in
perfect agreement with the older data of Betts et al. \cite{Be81b}. The fact
that the measured angular distributions correspond to shapes characterized by a
single Legendre polynomial squared means that the resonant behavior is
dominated by a unique and pure partial wave associated with the angular
momentum value L = 38$\hbar$. This value can finally be considered as the spin
of a well defined and isolated quasi-molecular resonance. It should be noticed
that this value is greater than the critical angular momentum that can be
extacted from the complete fusion data \cite{Di81,Vi90} L$_{crit}$ $\approx$
35$\hbar$ and lower than the grazing angular momentum L$_{graz}$ $\approx$
38-40$\hbar$ as obtained from a quarter point analysis of the elastic
scattering \cite{Di81}. The total cross section of the identical particle
exit-channel which can be estimated from the excitation energy spectra is found
to be around 10-12 mb, i.e. approximatively 1$\%$ of the total complete fusion
cross section \cite{Di81,Vi90}. FF calculations \cite{Fa96,Ma97} do predict
approximatively 5-10 mb for the symmetric-mass fission cross section. 

The fact that the angular momentum L = 38$\hbar$ is dominant in these resonant
channels means that the projection of the spin along the direction
perpendicular to the reaction plane is m=0. 

\newpage

\centerline{\bf IV. Fragment-fragment-$\gamma$ coincidence data} 

\vskip 1.0 cm

In this section we will focus our analysis of the fragment-fragment-$\gamma$
coincidence data on the $^{28}$Si+$^{28}$Si symmetric-fission exit-channel as
defined in the previous section. Doppler-shift corrections were applied to the
$\gamma$-ray data of fig.6 on an event-by-event basis using the measured
velocities of the detected $^{28}$Si fragments. Since it is not possible to
know {\it a priori} which of the two fragments emits the detected $\gamma$-ray,
both Doppler corrections were applied by using the method developed in
Ref.\cite{Fa96,Sa94}. In order to investigate the resonant effects on the
feeding of the $^{28}$Si states and to search for highly deformed bands in
$^{28}$Si, we have selected the 73.2$^{o}$ $\leq$ $\theta$$_{c.m.}$ $\leq$
105.6$^{o}$ angular region where the angular distributions strongly oscillate
as shown in fig.5. This region will be called in the following as the resonance
region because strong absorption features are known to be less significant due
to the Coulomb repulsion. 

The feeding of $^{28}$Si states in the resonance region, which are presented in
fig.7, indicates that the K$^{\pi}$ = 3$_{1}^{-}$ band is more strongly fed
than the 3$_{1}^{+}$ band. The population of the 6.88 MeV collective 3$^{-}$
state, which was not apparent in the fragment-fragment coincident data as
presented in fig.4, is quite strong. This is an indication that $^{28}$Si has
an {\bf \it oblate deformed} shape when the resonant features are present. 

As already mentionned in a study of the $^{24}$Mg($^{32}$S,$^{28}$Si)$^{28}$Si
reaction \cite{Sa94}, the population of the second excited band K$^{\pi}$ =
0$^{+}$ appears to be significantly well populated. Statistical-model
calculations \cite{Fa96,Sa94} are in progress in order to check whether the
role of this highly deformed prolate band in $^{28}$Si is of importance in the
resonant structure. 

One of the more particularly interesting feature in the ground state K$^{\pi}$
= 0$_{1}^{+}$ band is that the mutually excited states are more intensily fed
that the singly excited states (0$_{1}^{+}$,
0$_{1}^{+}$),(2$_{1}^{+}$,0$_{1}^{+}$), and (4$_{1}^{+}$,0$_{1}^{+}$) and the
relative ratios between them are equal to 12 $\%$. This high degree of
selectivity in the population of mutually excited yrast states in both
$^{28}$Si fragments definitively confirms the early findings of Betts et al.
\cite{Be81c}. 

The complexity of the higher energy structures apparent in fig.4 obtained with
the particle data makes it very difficult to assign specific mutual excitations
to the structures observed at energy above $\approx$ 6.5 MeV. This can be
however mostly resolved by the $\gamma$-ray data as shown by the two examples
displayed in fig.8. The tentative spin assignments of the mutual yrast states
as populated by the $^{28}$Si+$^{28}$Si symmetric-fission exit-channels with a
very high degree of selectivity are more firmly identified by the
fragment-fragment-$\gamma$ data. The particle spectra of fig.8 have been
obtained by gating on $\gamma$-rays of the 2$_{1779}^{+}$ $\rightarrow$
0$_{gs}^{+}$ and 4$_{4617}^{+}$ $\rightarrow$ 2$_{1779}^{+}$ transitions
respectively. It is evident that the populations of the mutual excitations are
dominant. However contributions from single excitation, such as the 4$^{+}$
state at 4.617 MeV which appears as a small shoulder on the (2$^{+}$,2$^{+}$)
peak \cite{Be81c,Sa94} of the ``inclusive" spectrum, are also present. This may
partly contradict the early suggestion that the resonant yields result primarly
from excitations of the yrast levels. 

We have been able to construct other ``coincident" spectra obtained with gates
on the weaker 3$_{6276}^{+}$ $\rightarrow$ 2$_{1779}^{+}$, 4$_{6888}^{+}$
$\rightarrow$ 2$_{1779}^{+}$ and 3$_{6276}^{+}$ $\rightarrow$ 0$_{gs}^{+}$
$\gamma$-ray transitions. They all show that non-negligeable contributions in
the so-called (2$^{+}$,2$^{+}$) and (4$^{+}$,2$^{+}$) peaks of fig.4 arise also
from the K$^{\pi}$ = 3$_{1}^{-}$ and 3$_{1}^{+}$ respectively. 

Spin-alignment estimations of the low-lying excitation states (single inelastic
2$^{+}$ and mutual inelastic (2$^{+}$,2$^{+}$) exit-channels) have been deduced
by measuring their particle-$\gamma$ angular correlations with EUROGAM Phase
II. Three quantization axes have been defined as follows: \\ 
a) the first axis corresponds to the beam axis,\\
b) the direction normal to the reaction plane represents the second axis,\\
c) and finally the molecular axis is parallel to the relative vector which
connects the two centers of the out-going binary fragments. 

The same features are found for the two exit channels. In fig.9 the results of
the $\gamma$-ray angular correlations for the mutual excitation exit-channel
are shown because of its symmetry. The minima observed in a) and b) at 90$^{o}$
imply that the intrinsic spin vectors of the 2$^{+}$ states lie in the reaction
plane and cannot be coupled with the relative orbital angular momentum and the
value of the total angular momentum remain J=38$\hbar$ for the two exit
channels in agreement with fig.5. The maximum around 90$^{o}$ in c) means that
the $^{28}$Si spin vectors are lying in the reaction plane with opposite
directions. The condition that the m=0 configuration is the dominant mode of
the collision in the $^{28}$Si+$^{28}$Si inelastic and mutual reactions
fulfills the classical rolling limit which demands parallel intrinsic spins
pointing in opposite directions. It can be pointed out that this condition
along with the strong population of the 3$^{-}$ oblate deformed band can be
relevant for the formation of a molecular di-nuclear complex with two
pancake-like $^{28}$Si nuclei touching each other edge to edge so that the spin
vectors of the fragments are in the pancakes' plane. 

The weak coupling evidenced in the present work for the $^{28}$Si+$^{28}$Si
system might be found to be in apparent contradiction with previous spin
alignment measurements as performed for the $^{24}$Mg+$^{24}$Mg system
\cite{Wu87,Ma87,Wu90}. However preliminary calculations \cite{Ue97} with the
molecular model developed by Uegaki and Abe \cite{Ue94,Ue89,Ue93} do predict an
oblate-oblate system with an equator-equator stable configuration (butterfly
mode) for the studied reaction, whereas the resonance-like structure observed
for $^{24}$Mg+$^{24}$Mg appears to be linked to a prolate-prolate system in a
pole-pole configuration (anti-butterfly mode). In the case of collisions
between two oblate deformed nuclei, there is some hints that the
equator-equator orientation is the most important one for molecular resonances
\cite{Ta78}. The observed mismatch of the spin vectors in $^{28}$Si+$^{28}$Si
with the orbital angular momentum might be discussed \cite{Ue97} as a candidate
of a butterfly mode excitation where the intrinsic spins of the two interacting
nuclei couple to zero. 

\newpage

\centerline{\bf V. Conclusions and Perspectives }

\vskip 1.0 cm

The resonant behavior of the $^{28}$Si+$^{28}$Si reaction at E$_{lab}$ = 111.6
MeV is clearly confirmed by the present fragment-fragment coincidence data for
the elastic and inelastic decay channels of single and mutual excitations. 
From the analysis of the particle angular distributions of these channels it
can be concluded that : \\ 
1) the J$^\pi$=38$^{+}$ resonance is now well defined, \\
2) the spin vectors of the $^{28}$Si fragments do not couple with the orbital
angular momentum: m=0.

The fragment-fragment-$\gamma$ coincident data demonstrate that for the
$^{28}$Si exit fragments : \\
1) the mutually excited states are the more strongly populated, \\ 
2) the population of the highly deformed prolate K$^\pi$ = 0$_{3}^{+}$ band
is enhanced, \\
3) the K$^\pi$ = 3$_{1}^{-}$ band is strongly fed indicative of an {\bf \it
oblate} deformed shape for the fragments,\\
4) the $^{28}$Si spin vectors are lying in the reaction plane in opposite
directions.

Qualitative arguments are in favor of the calculations of the molecular model
of Uegaki and Abe \cite{Ue97} which predict a butterfly mode excitation for the
$^{28}$Si+$^{28}$Si system, while an anti-butterfly motion is induced in the
$^{24}$Mg+$^{24}$Mg collision. A statistical-model calculation \cite{Fa96} is
in progress to describe both the structures observed at high excitation energy
in the fission Q-value spectra and the population pattern of states in the
$^{28}$Si fission fragments. This will suggest that these states may also
reflect nuclear structure effects at the point of fission. 

Similar studies are underway in the analysis of the
$^{28}$Si($^{28}$Si,$^{24}$Mg)$^{32}$S and \\
$^{28}$Si($^{28}$Si,$\alpha$$\alpha$)$^{48}$Cr exit-channels. This work call
for experiments with EUROBALL and/or GAMMASPHERE to measure precise excitation
functions of the $^{24}$Mg+$^{24}$Mg reaction which will also help in
developing the differences and possible relationships between the heavy-ion
resonance and compound-nucleus fission processes in light systems. 

\vskip 1.0cm

{\bf Acknowledgments :} The authors wish to thank Professor Y. Abe and
Professor E. Uegaki for their stimulating theoretical help in the initial
stages of this work. We would like also to acknowledge the VIVITRON staff
for providing us good $^{28}$Si beams. S. Szilner is warmly thanked for her
assistance during the data taking. One of us would like to acknowledge
Professor R.R. Betts for useful discussions on various aspects of this work and
for communicating us the data of Ref.\cite{Be97} before publication.

\newpage

\begin{figure}

FIGURE 1 : Cranking Nilsson-Strutinsky calculations of potential energy
surfaces for $^{56}$Ni at high spin I = 40$\hbar$ shown plotted as a function
of deformation $\beta$ and mass-asymmetry M$_{r}$ = M/56 where M is the mass of
one of the fragment (figure taken from \cite{Be85}). 

\end{figure}
\begin{figure}

FIGURE 2 : Angle-integrated yields of the $^{28}$Si+$^{28}$Si elastic
scattering, 2$^{+}$, mutual 2$^{+}$, and mutual (4$^{+}$,2$^{+}$) excitations
measured as a function of E$_{c.m.}$ in 100 keV steps (figure taken from 
\cite{Be81a}). The top part of the figure shows in the total yield of the
$^{28}$Si+$^{28}$Si exit-channel that many of the resonant structures of width
100-200 keV are well correlated. 

\end{figure}
\begin{figure}

FIGURE 3 : Ejectile energy E$_{3}$ versus excitation energy E$_{x}$
two-dimensional plot of the $^{28}$Si+$^{28}$Si symmetric-fission exit-channel. 

\end{figure}
\begin{figure}

FIGURE 4 : Excitation energy E$_{x}$ spectrum for the $^{28}$Si+$^{28}$Si exit
channel measured at large angles. Tentative assignments of the resolved and
non-resolved peaks are indicated. 

\end{figure}
\begin{figure}

FIGURE 5 : Angular distributions of the elastic, inelastic 2$_{1}^{+}$ and
mutual (2$_{1}^{+}$,2$_{1}^{1}$), (4$_{1}^{+}$,0$_{1}^{+}$) excitation
channels. The solid lines are [P$_{L}$(cos$\theta$)]$^{2}$ shapes for L =
38$\hbar$. 

\end{figure}
\begin{figure}

FIGURE 6 : $\gamma$-ray spectrum in coincidence with the $^{28}$Si+$^{28}$Si
symmetric-fission exit-channel.

\end{figure}
\begin{figure}

FIGURE 7 : Measured feedings for states in $^{28}$Si populated in the resonance
region. 

\end{figure}
\begin{figure}

FIGURE 8 : Excitation-energy spectra of the $^{28}$Si+$^{28}$Si exit-channel as
gated by the 2$_{1}^{+}$ $\rightarrow$ gs and 4$_{1}^{+}$ $\rightarrow$
2$_{1}^{+}$ $\gamma$-ray transitions.

\end{figure}
\begin{figure}

FIGURE 9 : $\gamma$-ray angular correlations of the (2$_{1}^{+}$,2$_{1}^{+}$)
states of the $^{28}$Si+$^{28}$Si exit-channel for 3 different quantization
axes. 

\end{figure}


\newpage

\begin{references}

\bibitem{Fa96} K.A. Farrar et al., Phys. Rev. C {\bf 54}, 1249 (1996).

\bibitem{Be96} C. Beck et al., Phys. Rev. C {\bf 54}, 227 (1996).

\bibitem{No96} R. Nouicer et al., Zeit. f\"ur Phys. {\bf A356}, 5 (1996). 

\bibitem{Ma97} T. Matsuse et al., Phys. Rev. C {\bf 55 }, 1380 (1997).

\bibitem{Be81a} R.R. Betts et al., Phys. Rev. Lett. {\bf 47}, 23 (1981).

\bibitem{Be95} C. Beck et al., Nucl. Phys. {\bf A583}, 269 (1995).

\bibitem{Be94} C. Beck et al., Phys. Rev. C {\bf 49}, 2618 (1994).

\bibitem{Be85} R.R. Betts et al., Nucl. Phys. {\bf A447}, 257 (1985).

\bibitem{Be81b} R.R. Betts et al., Phys. Lett. {\bf 100B}, 117 (1981).

\bibitem{Ra92} W.D.M. Rae et al., Phys. Lett. {\bf B279}, 207 (1992).

\bibitem{Ue94} E. Uegaki and Y. Abe, Phys. Lett. {\bf B340}, 143 (1994).

\bibitem{Ue89} E. Uegaki and Y. Abe, Phys. Lett. {\bf B231}, 28 (1989).

\bibitem{Ue93} E. Uegaki and Y. Abe, Prog. Theor. Phys. {\bf 90}, 615 (1993).

\bibitem{Be81c} R.R. Betts et al., Phys. Rev. Lett. {\bf 46}, 313 (1981).

\bibitem{Be97} R.R. Betts, private communication of unpublished data.

\bibitem{No97} R. Nouicer, Ph.D. Thesis, Universit\'e Louis Pasteur, Strasbourg
(unpublished).

\bibitem{Di81} S.B. DiCenzo et al., Phys. Rev. C {\bf 23}, 2561 (1981); S.B.
DiCenzo, Ph.D. Thesis, University of Yale, 1980 (unpublished). 

\bibitem{Vi90} M. Vineyard et al., Phys. Rev. C {\bf 41}, 1005 (1990) and
references therein.

\bibitem{Sa94} S.J. Sanders et al., Phys. Rev. C {\bf 49}, 1016 (1994).

\bibitem{Wu87} A.H. Wuosmaa et al., Phys. Rev. Lett. {\bf 58}, 1312 (1987).

\bibitem{Ma87} A. Mattis et al., Phys. Lett. {\bf B191},, 328 (1987).

\bibitem{Wu90} A.H. Wuosmaa et al., Phys. Rev. C {\bf 41}, 2666 (1990).

\bibitem{Ue97} Y. Abe and E. Uegaki, private communications.

\bibitem{Ta78} O. Tanimura and T. Tazawa, Phys. Lett. {\bf 78B}, 1 (1978).

\end{references}
\end{document}